\documentclass[journal,onecolumn,12pt]{IEEEtran}
\renewcommand{\baselinestretch}{1.5}

\usepackage{epsfig,amsmath,amsthm,amssymb,epsf,algorithm,algpseudocode}
\usepackage{graphicx}
\usepackage{subfig}
\usepackage{array}
\usepackage{float}
\usepackage{booktabs}
\usepackage{multirow}
\usepackage{color}

\newtheorem{definition}{Definition}
\newtheorem*{rem}{Remark}
\newtheorem{exm}{Example}

\newcounter{mytempeqncnt}

\def\b0{{\pmb{0}}}

   \def\bv{{\mathbf{v}}}
\def\bc{{\mathbf{c}}}   
   \def\bx{{\mathbf{x}}}
   \def\by{{\mathbf{y}}}

 \def\bH{{\mathbf{H}}}  
  \def\bP{{\mathbf{P}}} \def\bV{{\mathbf{V}}}

\def\spp{{\mathfrak{p}}}

\begin{document}
%
\title{Implementation of Decoders for LDPC \\
Block Codes and LDPC Convolutional \\ Codes Based on GPUs}
%
%
%

\author{Yue~Zhao and 
        Francis~C.M.~Lau,~\IEEEmembership{Senior~Member,~IEEE}
\IEEEcompsocitemizethanks{\IEEEcompsocthanksitem Yue Zhao and Francis Lau are with 
the Department of Electronic and Information Engineering, 
The Hong Kong Polytechnic University.}
}

%
%

%


\IEEEcompsoctitleabstractindextext{
\begin{abstract}
With the use of belief propagation (BP) decoding algorithm, low-density
parity-check (LDPC) codes can achieve near-Shannon limit performance. LDPC codes
can accomplish bit error rates (BERs) as low as $10^{-15}$ even at a small
bit-energy-to-noise-power-spectral-density ratio ($E_{b}/N_{0}$). In order to
evaluate the error performance of LDPC codes, simulators running on central
processing units (CPUs) are commonly used. However, the time taken to evaluate
LDPC codes with very good error performance is excessive. For example, assuming
$30$ iterations are used in the decoder, our simulation results have shown that
it takes a modern CPU more than 7 days to arrive at a BER of $10^{-6}$ for a code
with length $18360$. In this paper, efficient LDPC block-code decoders/simulators 
which run on graphics processing units (GPUs) are proposed. 
We also implement the decoder for the LDPC convolutional code (LDPCCC). The
LDPCCC is derived from a pre-designed quasi-cyclic LDPC block code with good
error performance. Compared to the decoder based on the randomly constructed
LDPCCC code, the complexity of the proposed LDPCCC decoder is reduced due to the
periodicity of the derived LDPCCC and the properties of the quasi-cyclic
structure.
{\color{black}
In our proposed decoder architecture, $\Gamma$ ($\Gamma$ is a multiple of a warp)
codewords are decoded together and hence the messages of $\Gamma$ codewords are
also processed together. 
Since all the $\Gamma$ codewords share the same Tanner graph,
messages of the $\Gamma$ distinct codewords corresponding to the same edge can be
grouped into one package and stored linearly. 
}
By optimizing the data structures of the messages used in the decoding process, 
both the read \emph{and} write processes can be performed in a highly parallel 
manner by the GPUs. In addition, a thread hierarchy 
{\color{black} minimizing 
}
the divergence of 
the threads is deployed, and it can maximize the efficiency of the parallel 
execution. 
{\color{black}
With the use of a large number of cores in the GPU to perform the 
simple computations simultaneously, our GPU-based LDPC decoder can obtain hundreds 
of times speedup compared with a 
serial CPU-based simulator and over $40$ times speedup compared with an $8$-thread CPU-based simulator.
}
\end{abstract}

}

\maketitle


\section{Introduction}
\label{sec:intro}

\IEEEPARstart{L}{ow-density} parity-check (LDPC) codes were invented by Robert Gallager
\cite{gallager_low-density_1963} but had been ignored for years until Mackay
rediscovered them \cite{mackay_good_1999}. They have attracted much attention
recently because they can achieve excellent error 
{\color{black} correcting }
performance based on the belief
propagation (BP) decoding algorithm. 


However, the BP decoding algorithm requires intensive computations. 
For applications like optical communication
\cite{djordjevic_using_2007,miyata_triple-concatenated_2010} which requires BERs
down to $10^{-15}$, using CPU-based programs to simulate the LDPC decoder 
{\color{black} is impractical. }
 Fortunately, the decoding algorithm possesses a high
data-parallelism feature, i.e., the data used in the decoding process are
manipulated in a very similar manner and can be processed separately from one
another. Thus, practical decoders with low-latency and high-throughput can be
implemented with dedicated hardware such as field programmable gate arrays
(FPGAs) or application specific integrated circuits (ASICs)
\cite{yanni_chen_fpga_2003,djordjevic_next_2009,benjamin_levine_implementation_2000,
pusane_implementation_2008,SBates08,CBK08,Swamy2005}.
However, high performance FPGAs and ASICs are very expensive and are
non-affordable by most researchers. Such hardware solutions also cost a long time
to develop. In addition, the hardware control and interconnection frame are
always associated with a specific LDPC code. If one parameter of an LDPC
code/decoder changes, the corresponding hardware design has to be changed
accordingly, rendering the hardware-based solutions non-flexible and
non-scalable.

Recently, graphics processing units (GPUs) used to process graphics only have
been applied to support general purpose computations \cite{falcao_how_2009}. In
fact, GPUs are highly parallel structures with many processing units. They
support floating point arithmetics and can hence conduct computations with the
same precision as CPUs. GPUs are particularly efficient in carrying out the same
operations to a large amount of (different) data. Compared with modern CPUs, GPUs
can also provide much higher data-parallelism and bandwidth. Consequently, GPUs
can provide a cheap, flexible and efficient solution of simulating an LDPC
decoder. Potentially, the simulation time can be reduced from 
{\color{black} months to weeks or days}
 when GPUs, instead of CPUs, are used. In addition, the GPU programming
codes can be re-used without much modification should more advanced GPUs 
{\color{black} be}
produced by manufacturers.

In \cite{ji_massively_2009,springerlink:10.1007}, a compressed parity-check
matrix has been proposed to store the indices of the passing messages in a cyclic
or quasi-cyclic LDPC code. Further, the matrix is stored in the constant cache
memory on the GPU for fast access. The messages are stored in a compressed manner
such that the global memory can be accessed in a coalesced way frequently.
However, the coalesced memory access occurs only during the data-read process and
is not always guaranteed due to a lack of data alignment. In
\cite{falcao_how_2009,falcao_massive_2008,falcao_massively_2011}, the sum-product LDPC decoder and the
min-sum decoder have been implemented with GPUs. Moreover, by combining sixteen
fixed-point $8$-bit data to form one $128$-bit data, the LDPC decoder in
\cite{falcao_how_2009} decodes sixteen codewords simultaneously and achieves a
high throughput. Although the method in \cite{falcao_how_2009} allows
coalesced memory access in \textit{either} the read \textit{or} write process,
coalesced memory access in \textit{both} the read \textit{and} write processes is
yet to be achieved.

Furthermore, the LDPC convolutional codes (LDPCCCs), first proposed in
\cite{jimenez_felstrom_time-varying_1999}, have been shown to achieve a better
error performance than the LDPC block code counterpart of similar decoding
complexity. There are many features of LDPCCC that make it suitable for real
applications. First, the LDPCCC inherits the structure of the convolutional code,
which allows continuous encoding and decoding of variable-length codes. Thus the
transmission of codewords with varying code length is possible. Second, the
LDPCCC adopts a pipelined decoding architecture --- in the iterative decoding
procedure, each iteration is processed by a separate processor and the procedure
can be performed in parallel. So a high-throughput decoder architecture is
possible. In \cite{TMKF08,MTBF07}, the concepts and realization of highly
parallelized decoder architectures have been presented and discussed. 
To the author's best knowledge, there is
not any GPU-based implementation of the LDPCCC decoder yet. The reason may lie in
the complexity structure of the LDPCCC compared to the LDPC block code,
particularly the random time-varying LDPCCC.

As will be discussed in this paper, an LDPCCC derived from a well designed QC-LDPC
code possesses not only the good BER performance, but also the regular structure
that results in many advantages in practical implementations. Due to the
structure inherited from the QC-LDPC code, the LDPCCC decoder enables an
efficient and compact memory storage of the messages with a simple address
controller.


In this paper, we develop flexible and highly parallel GPU-based decoders for the
LDPC codes. 
We improve the efficiency by making
(i) the threads of a warp follow the same execution path 
{\color{black} (except when
deciding whether a bit is a ``0'' or a ``1'') 
}
and (ii) the memory
accessed by a warp be of a certain size and be aligned. 
The results show that the decoders based on the GPUs achieve 
remarkable speed-up improvement --- more than $100$ times faster than the
{\color{black} serial}
 CPU-based
decoder.

We also develop a GPU-based decoder for the LDPC convolutional codes. 
We propose a decoder architecture
for LDPCCC derived from QC-LDPC block-code. By taking advantage of the
homogeneous operations of the pipeline processors, we compress the index
information of different processors into one lookup table. Combined with an
efficient thread layout, the decoder is optimized in terms of thread execution
and memory access. 
Simulation results show that compared with
the 
{\color{black} serial}
 CPU-based decoder, the GPU-based one can achieve as many as $200$ times
speed-up.
{\color{black} The GPU-based decoder, moreover, outperforms a quad-core CPU-based
decoder by almost $40$ times in terms of simulation time.}

The rest of the paper is organized as follows. Section~\ref{sec:review} 
reviews the structure and decoding algorithm of the LDPC code. The same
section also reviews the construction of LDPCCC based on QC-LDPC code as well as the
decoding process for the LDPCCC. In Section~\ref{sec:cuda}, the architecture of
CUDA GPU and the CUDA programming model is introduced. Section~\ref{sec:dec}
describes the implementation of the LDPC decoder and LDPCCC decoder based on
GPUs. Section~\ref{sec:results} presents the simulation results of the LDPC
decoder and LDPCCC decoder. 
{\color{black} The decoding times
 are compared when (i) a GPU is used, (ii) a quad-core CPU
is used with a single thread, and
(iii) a quad-core CPU
is used with up to 8 threads. }
Finally, Section~\ref{sec:conclude}
concludes the paper.

\section{Review of LDPC Codes and LDPC Convolutional Codes}
\label{sec:review}

\subsection{Structure of LDPC Codes and QC-LDPC Codes}

A binary $(N,K)$ LDPC code is a linear block code specified by a sparse $M\times
N$ parity-check matrix $\bH$, where $M=N-K$. The code rate of such an LDPC code
is $R\ge K/N=1-M/N$. The equality holds when $\mathbf{H}$ is full rank.

The $\bH$ matrix contains mostly $0's$ and relatively a small number of $1's.$
Such a sparsity structure is the key characteristic that guarantees good
performance of LDPC codes. A \textit{regular} LDPC code is a linear block code
with $\mathbf{H}$ containing a constant number $w_{c}$ of $1$'s in each column
and a constant number $w_{r}$ of $1$'s in each row. Moreover, $w_{r}$ and $w_{c}$
satisfy the equation $w_{r}=w_{c}\times\frac{N}{M}$. Otherwise the code is
defined as an \emph{irregular} LDPC code.

\begin{exm} 
The parity-check matrix \textbf{$\mathbf{H}$} in (\ref{eq:ldpcc})
shows an example of an irregular LDPC code.
\begin{equation}
\mathbf{H}=
\begin{bmatrix}
1 & 1 & 1 & 1 & 0 & 1 & 0 & 1 & 0 & 1\\
1 & 0 & 1 & 1 & 0 & 0 & 1 & 1 & 1 & 1\\
0 & 1 & 0 & 1 & 0 & 0 & 0 & 1 & 0 & 0\\
1 & 0 & 0 & 0 & 1 & 0 & 1 & 1 & 1 & 1\\
0 & 0 & 1 & 1 & 1 & 0 & 1 & 0 & 1 & 0
\end{bmatrix}
\label{eq:ldpcc}
\end{equation}
\end{exm}

\begin{figure}[t]
\centering
\includegraphics[width=0.7\columnwidth]{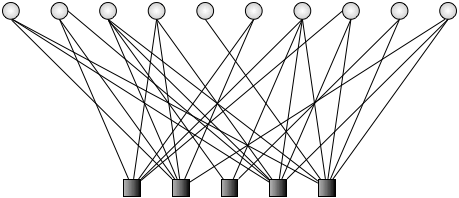}
\caption{Tanner Graph representation of the LDPC code defined by \eqref{eq:ldpcc}.}
\label{Flo:tannerldpc}
\end{figure}

A bipartite graph called Tanner graph \cite{tanner_recursive_1981} can be used to
represent the codes and to visualize the message-passing algorithm.
Figure~\ref{Flo:tannerldpc} is the underlying Tanner graph of the $\mathbf{H}$ in
\eqref{eq:ldpcc}. The $N$  
{\color{black}  upper nodes}
 are called the message nodes or the
variable nodes and the $M$  
{\color{black} nodes in the lower part of Fig.~\ref{Flo:tannerldpc} }
are called the check nodes. 
An edge in the Tanner graph represents the adjacency
of the variable node $i$ and the check node $j$. It corresponds to a nonzero
$(i,j)$-th entry in the $\mathbf{H}$ matrix.

QC-LDPC codes form a subclass of LDPC codes with the parity-check matrix consisting
of circulant permutation matrices \cite{fossorier_quasicyclic_2004,Tam:2010lr}.
The parity-check matrix of a regular $(J, L)$ QC-LDPC code is represented by
\begin{equation} 
\bH = 
\begin{bmatrix}
\bP^{a_{1,1}} & \bP^{a_{1,2}} & \cdots & \bP^{a_{1,L}} \\
\bP^{a_{2,1}} & \bP^{a_{2,2}} & \cdots & \bP^{a_{2,L}} \\
\vdots & \cdots & \cdots & \vdots  \\
\bP^{a_{J,1}}  & \bP^{a_{J,2}} & \cdots & \bP^{a_{J,L}} 
\end{bmatrix},
\label{eq:qcH}
\end{equation}
where $J$ denotes the number of block rows, $L$ is the number of block columns,
$\bP$ is the identity matrix of size $p \times p$, and $\bP^{a_{j,l}}$ ($1 \leq j
\leq J$; $1 \leq l \leq L$) is a circulant matrix formed by shifting the columns
of $\bP$ cyclically to the right $a_{j,l}$ times with $a_{j,l}$'s being
non-negative integers less than $p$. The code rate $R$ of $\bH$ is lower bounded
by $R \geq 1- J / L$. If one or more of the sub-matrix(matrices) is/are
substituted by the zero matrix rendering non-uniform distributions of the
check-node degrees or variable-node degrees, the QC-LDPC code becomes an
irregular code.

\subsection{Belief Propagation Decoding Algorithm for LPDC Codes}
\label{sec:ldpc_dec}

LDPC codes are most commonly decoded using the belief propagation (BP) algorithm
\cite{richardson_design_2001,richardson_efficient_2001}. Referring to the Tanner
graph shown in Fig.~\ref{Flo:tannerldpc}, the variable nodes and the check nodes
exchange soft messages iteratively based on the connections and according to a
two-phase schedule. 

Given a binary (\emph{N, K}) LDPC code with a parity-check matrix $\mathbf{H}$,
we define $\mathcal{C}$ as the set of binary codewords $\mathbf{c}$ that satisfy
the equation $\mathbf{c}\mathbf{H}^{\mathrm{T}}=\mathbf{0}$. At the transmitter
side, a binary codeword $\mathbf{c}=(c_{0},c_{1},\ldots,c_{N-1})$ is mapped into
the sequence $\mathbf{x}=(x_{0},x_{1},\ldots,x_{N-1})$ according to
$x_{n}=1-2c_{n}$. We assume that $\mathbf{x}$ is then transmitted over an
additive white Gaussian noise (AWGN) channel and the received signal vector is
then given by $\mathbf{y}=(y_{0},y_{1},\ldots,y_{N-1})=\mathbf{x}+\mathbf{g}$,
where $\mathbf{g}=(g_{0},g_{1},\ldots,g_{N-1})$ consists of independent Gaussian
random variables with zero mean and variance $\sigma^{2}=N_{0}/2$.

Let $\mu_{n}$ be the initial log-likelihood ratio (LLR) that the variable node
$n$ is a ``$0$'' to that it is a ``$1$'', i.e.,
\begin{equation}
\mu_{n}=\ln\left(\frac{\Pr(c_{n}=0|y_{n})}{\Pr(c_{n}=1|y_{n})}\right).
\end{equation}
Initially, $\mu_{n}$ is calculated by $\mu_{n}=(4/N_{0})\cdot
y_{n}=\frac{2y_n}{\sigma^2}$ \cite{xiao-yu_hu_efficient_2001}. Define
$\mathcal{N}(m)$ as the set of variable nodes that participate in check node $m$
and $\mathcal{M}(n)$ as the set of check nodes connected to variable node $n$. At
iteration $i$, let $\beta_{mn}^{(i)}$ be the LLR messages passed from variable
node $n$ to check node $m$; $\alpha_{mn}^{(i)}$ be the LLR messages passed from
check node $m$ to variable node $n$; and $\beta_{n}^{(i)}$ be the \emph{a
posteriori} LLR of variable node $n$. Then the standard BP algorithm can be
described in Algorithm~\ref{alg:ldpc_dec}~\cite{mackay_good_1999,chen_reduced-complexity_2005}.

{\color{black} 
Note that the decoding algorithm consists of 4 main procedures: initialization, horizontal step,
vertical step and making hard decisions. For each of these procedures, multiple threads can be used in executing the computations in parallel and all the threads will follow the same instructions with no divergence occurring, except when making hard decisions.
}


%
\begin{algorithm}[ht]
\caption{\color{black} BP Decoding Algorithm for LDPC Code}
\label{alg:ldpc_dec}
\begin{algorithmic}[1]                    

{\color{black} 
\Statex [Initialization]
	\For{$0\le n\le N-1$ and $m\in\mathcal{M}(n)$} 
		\State $\beta_{mn}^{(0)}=\mu_{n}$
	\EndFor
	\State Reset the iteration counter $i=1$

\Statex [Horizontal step]
\label{alg:line:horizontal}
	\For{$0\le m\le M-1$ and $n\in\mathcal{N}(m)$}
		\State Update the check-to-variable messages by
		\begin{equation}
		\alpha_{mn}^{(i)}=2\tanh^{-1}\Bigg(\prod_{n\prime\in\mathcal{N}(m)\backslash n}\tanh\left(\frac{\beta_{mn\prime}^{(i-1)}}{2}\right)\Bigg)
		\label{eq:bphs}
		\end{equation}
		where $\mathcal{N}(m)\backslash n$ denotes the set $\mathcal{N}(m)$
		excluding the variable node $n$
	\EndFor
	
\Statex [Vertical step]
	\For{$0\le n\le N-1$ and $m\in\mathcal{M}(n)$}
		\State Update the variable-to-check messages by
		\begin{equation}
		\beta_{mn}^{(i)}=\mu_{n}+\sum_{m\prime\in\mathcal{M}(n)\backslash m}\alpha_{m\prime n}^{(i)}
		\label{eq:bpvs}
		\end{equation}
		where $\mathcal{M}(n)\backslash m$ denotes the set $\mathcal{M}(n)$
		with check node $m$ excluded. 
		\State Calculate the \emph{a posteriori} LLRs using 
		\begin{equation}
		\beta_{n}^{(i)}=\mu_{n}+\sum_{m'\in\mathcal{M}(n)}\alpha_{m'n}^{(i)}
		\label{eq:bpd}
		\end{equation}
	\EndFor
}
\Statex {\color{black}  [Next iteration] 
	\If {$i < I_{max}$ (the max. no. of iterations)} 
		\State $i = i + 1$
		\State Go to the [Horizontal Step]	
	\EndIf 
}
\Statex   [Making hard decision]
	\State Make the hard decisions based on the LLRs 
	\begin{eqnarray}
	\hat{c}_{n}^{(i)} & = & \begin{cases}
	0 & {\rm if}\ \beta_{n}^{(i)}\ge0\\
	1 & {\rm if}\ \beta_{n}^{(i)}<0
	\end{cases}
	\label{eq:hrddcs}
	\end{eqnarray} 

\end{algorithmic}
\end{algorithm}

\subsection{Structure of LDPC Convolutional Codes}
\label{sec:ldpccc_intro}

A (time-varying) semi-infinite LDPC convolutional code can be represented
by its parity check matrix in \eqref{eq:ldpcccH}.
\begin{figure*}[t]
\normalsize
\setcounter{mytempeqncnt}{\value{equation}}
\setcounter{equation}{2}
\begin{equation}
\bH_{[0,\infty]}=
\begin{bmatrix}
\bH_{0}(0) & & & & & & \\
\bH_{1}(1) & \bH_{0}(1) & & & & & \\
\vdots  & \vdots  & \ddots & & & & \\
\bH_{m_s}(m_s) & \bH_{m_s-1}(m_s) & \cdots & \bH_{0}(m_s) & & & \\
 & \bH_{m_s}(m_s+1) & \bH_{m_s-1}(m_s+1) & \cdots & \bH_{0}(m_s+1) & & \\
 & & \ddots & & & \ddots & \\
 & & \bH_{m_s}(t) & \bH_{m_s-1}(t) & \cdots & \bH_{0}(t) & \\
 & & \ddots & \ddots & & & \ddots
\end{bmatrix},
\label{eq:ldpcccH}
\end{equation} 
\setcounter{equation}{\value{mytempeqncnt}}
\hrulefill
\vspace*{4pt}
\end{figure*}
where $m_s$ is referred to as the syndrome former memory of the 
parity-check matrix.
Besides, the sub-matrices $\bH_{i}(t), i = 0, 1, ...,m_s$, are binary
$(c - b) \times c$ matrices given by
\[
\bH_i(t) = 
\begin{bmatrix}
h_i^{(1,1)}(t) & \cdots & h_i^{(1,c)}(t) \\
\vdots & \ddots & \vdots \\
\vdots & \ddots & \vdots \\
h_i^{(c-b,1)}(t) & \cdots & h_i^{(c-b,c)}(t)
\end{bmatrix}.
\]
If $\bH_i(t)$ are full rank for all time instant $t$, the matrix $\bH$ in \eqref{eq:ldpcccH}
defines a rate $R = b/c$ convolutional code ignoring the irregularity at
the beginning.
\begin{definition}
A LDPC convolutional code is called a regular $(m_s,J,K)$-LDPC convolutional code 
if the parity-check matrix $\bH_{[0,\infty]}$ has exactly $K$ ones in
each row and $J$ ones in each column starting from the $(m_s\cdot(c-b)+1)$-th
row and $(m_s\cdot c + 1)$-th column.
\end{definition}
\begin{definition}
An $(m_s,J,K)$-LDPC convolutional
code is periodic with period $T$ if $\bH_i(t), i\in \mathbb{Z}^{+}$ is periodic, 
i.e., $\bH_i(t) = \bH_i(t+T), \forall i, t$. 
\end{definition}
%

A code sequence $\bv_{[0,\infty]}$ = $[\bv_0, \bv_1,...,\bv_{\infty}]$ 
is ``valid'' if it satisfies the equation
\begin{equation}
\bv_{[0,\infty]}\bH_{[0,\infty]}^T=\b0
\end{equation}
where $\bv_i$ = $(v_i^{(1)}, v_i^{(2)} ,..., v_i^{(c)})$ and
$\bH_{[0,\infty]}^T$ is the syndrome-former (transposed parity-check) 
matrix of $\bH_{[0,\infty]}$.


\subsection{Deriving LDPC Convolutional codes from QC-LDPC block codes}
\label{sec:ldpccc_derive}

There are several methods to construct LDPC convolutional codes from LDPC block
codes. One method is to derive time-varying LDPCCC by unwrapping randomly
constructed LDPC block codes \cite{jimenez_felstrom_time-varying_1999} and
another is by unwrapping the QC-LDPC codes \cite{tanner_ldpc_2004,Pusane2007}.
We now consider a construction method by unwrapping a class of QC-LDPC block 
code.

Suppose we have created a $(J, L)$ QC-LDPC block code $\bH_{QC}$ with $J$
block-rows and $L$ block-columns. The size of its circulant matrices is $p\times
p$. We can derive the parity-check matrix for a LDPC convolutional code using the
following steps.

\begin{enumerate}
\item Partition the $pJ\times pL$ parity-check matrix $\bH_{QC}$ to form 
a $\Lambda\times\Lambda$ matrix,
where $\Lambda$ is the greatest common divisor of $J$ and $L$, i.e.,
\[
\bH_{QC} = 
\begin{bmatrix}
\bH_{1,1} & \bH_{1,2} & \cdots & \bH_{1,\Lambda} \\
\bH_{2,1} & \bH_{2,2} & \cdots & \bH_{2,\Lambda} \\
\vdots & \vdots & \cdots & \vdots \\
\bH_{\Lambda,1} & \bH_{\Lambda,2} & \cdots & \bH_{\Lambda,\Lambda}
\end{bmatrix}_{\Lambda\times\Lambda},
\]
where $\bH_{i,j}$ is a $(pJ/\Lambda)\times(pL/\Lambda)$ matrix, 
for $i,j=1,2,...,\Lambda$.

\item Divide $\bH_{QC}$ along the diagonal into two portions: the 
strictly upper-triangular portion $\bH_{QC}^{(U)}$ and the 
lower-triangular portion $\bH_{QC}^{(L)}$ as follows:
\[
\bH_{QC}^{(U)} = 
\begin{bmatrix}
\mathbf{0} & \bH_{1,2} & \bH_{1,3} & \cdots & \bH_{1,\Lambda} \\
 & \mathbf{0} & \bH_{2,3} & \cdots & \bH_{2,\Lambda} \\
 & & \ddots & \cdots \vdots \\
 & & & \mathbf{0} & \bH_{\Lambda-1,\Lambda} \\
 & & & & \mathbf{0}
\end{bmatrix}_{\Lambda\times\Lambda},
\]
and
\[
\bH_{QC}^{(L)} = 
\begin{bmatrix}
\bH_{1,1} &   &   &  \\
\bH_{2,1} & \bH_{2,2} &   &  \\
\vdots & \vdots & \ddots &  \\
\bH_{\Lambda,1} & \bH_{\Lambda,2} & \cdots & \bH_{\Lambda,\Lambda}
\end{bmatrix}_{\Lambda\times\Lambda}.
\]
\item Unwrap the parity-check matrix of the block code to obtain the
parity-check matrix of LPDCCC. First paste the strictly upper-triangular
portion below the lower-triangular portion. Then repeat the resulting
diagonally-shaped matrix infinitely, i.e.,
\[
\bH_{conv} = 
\begin{bmatrix}
\bH_{QC}^{L} & & & \\
\bH_{QC}^{U} & \bH_{QC}^{L} & & \\
 & \bH_{QC}^{U} & \bH_{QC}^{L} & \\
 & & \ddots & \ddots \\
\end{bmatrix}.
\]
\end{enumerate}

The resulting time-varying LDPCCC has a period of $T = \Lambda$ and the memory
$m_s$ equals $\Lambda-1$. The girth of the
derived LPDCCC is at least as large as the girth of the QC-LDPC code
\cite{Lentmaier2010a}. A convenient feature of this time-varying unwrapping is
that a family of LDPC convolutional codes can be derived by choosing different
circulant size $p$ of the QC-LDPC block code.

\begin{figure*}[t]
\centering
\includegraphics[width=0.7\textwidth]{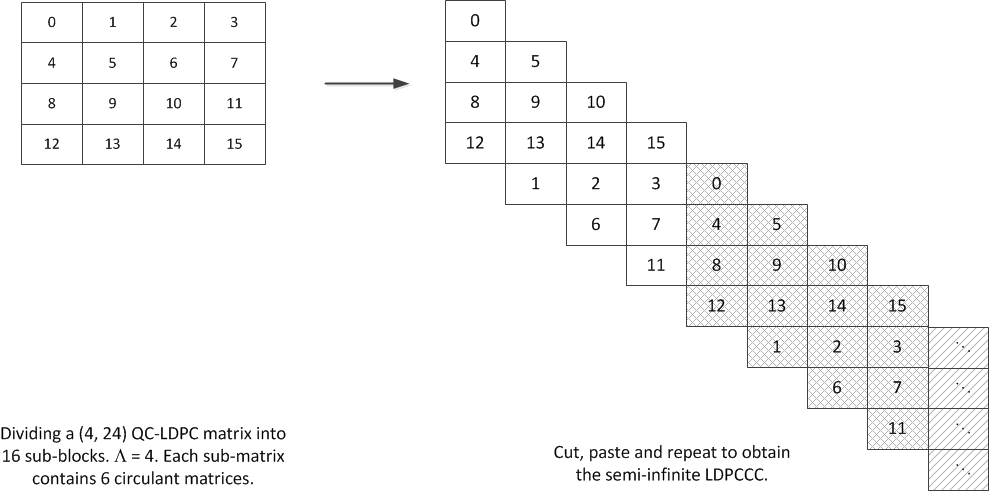}
\caption{Illustration of constructing a LPDCCC from a QC-LDPC block code.}
\label{fig:ldpccc_const}
\end{figure*}

\begin{exm}
Consider a QC-LDPC code with $4$ block rows and $24$ block columns, i.e., $J=4$ and $L=24$. 
It is first divided into $4\times4$ equally sized
sub-blocks\footnote{Here we use sub-block to denote the $(pJ/\Lambda)\times(pL/\Lambda)$ matrix as to
distinguish it with the sub-matrix within it, i.e., the $p\times p$ matrix.}, 
i.e., $\Lambda = 4$. Then the parity-check matrix of LDPCCC 
is derived. The construction process is shown in Fig.~\ref{fig:ldpccc_const}.
\label{exm:deriving_LDPCCC}
\end{exm}

\subsection{Decoding Algorithm for LDPCCC}
\label{sec:ldpccc_dec}

In $\bH_{[0,\infty]}$, two different variable nodes connected to the same check
node cannot be distant from each other more than $m_s$ time units. This allows a
decoding window that operates on a fixed number of nodes at one time. Since any
two variable nodes that are at least $m_s + 1$ units apart can be decoded
independently, parallel implementation is feasible. The LDPCCC can therefore be
decoded with pipelined BP decoding algorithm \cite{jimenez_felstrom_time-varying_1999}. 
Specifically, for a maximum iteration
number of $I$, $I$ independent processors will be employed working on different
variable nodes corresponding to different time.
{\color{black} 
In each processor, the variable nodes and the check nodes
exchange soft messages iteratively based on the connections and according to a
two-phase schedule. 
}
\begin{figure*}[t]
\centering
\includegraphics[width=\textwidth]{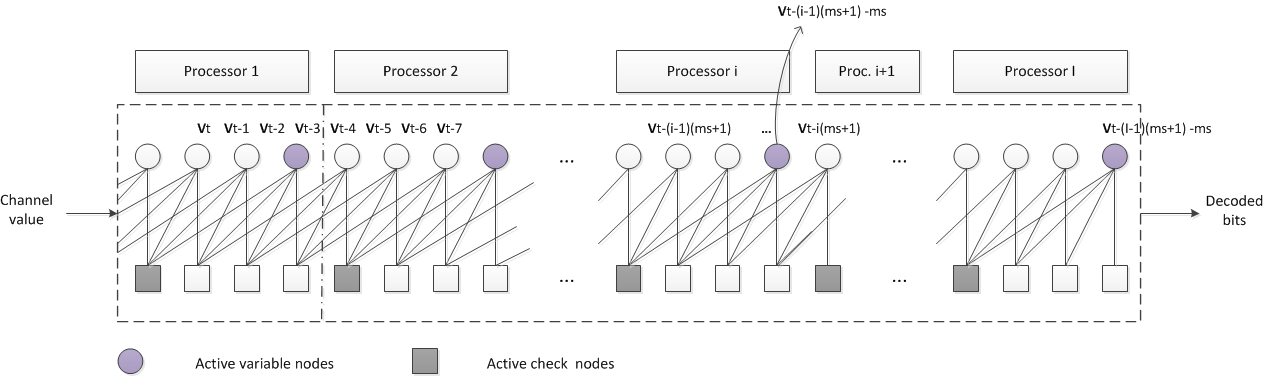}
\caption{Continuous decoding of LDPC convolutional code with $I$ processors. Each circle
denotes a group of $c$ variable nodes and each square denotes a group of $(c-b)$ check 
nodes. Each edge represents the connection between the $c$ variable
node and the $(c-b)$ check nodes.}
\label{fig:ldpcccdec}
\end{figure*}

Fig.~\ref{fig:ldpcccdec} shows a decoder on the Tanner graph. It is 
based on the LDPCCC structure shown in Example \ref{exm:deriving_LDPCCC}. 
The code has a rate of $R = 5/6$ and a syndrome
former memory of $m_s = 3$. We refer the $c$ incoming variable nodes (bits) as a
frame. Note that every $c$ bits form a frame and every $m_s + 1$ frames are
involved in the same constraints. 
The $I$ processors can
operate concurrently. At every iteration, every processor first updates the $(c -
b)$ neighboring check nodes of the $c$ variable nodes that just come into this
processor. Then every processor will update the $c$ variables which are
leaving this processor. 

The computations of the check-node updating and variable-node updating are based
on the standard BP algorithm 
Suppose $\bv_{[0,\infty]}$ =
$[\bv_0, \bv_1,\ldots,\bv_{\infty}]$, where $\bv_t = (v_t^{(1)}, v_t^{(2)} ,\ldots,
v_t^{(c)})$ is the $t$th transmitted codeword.
{\color{black}
Then the codeword $\bv_{[0,\infty]}$ is mapped into the sequence $\bx_{[0,\infty]}= [\bx_0, \bx_1,\ldots,\bx_{\infty}]$ according to $\bx_t = (x_t^{(1)}, x_t^{(2)} ,\ldots,
x_t^{(c)})$ and  $x_t^{(j)} = 1 - 2v_{t}^{(j)}$ ($j=1,2,\ldots,c$). Assuming 
 an AWGN channel, the received signal $\by_{[0,\infty]}= [\by_0, \by_1,\ldots,\by_{\infty}]$ is further given by 
$\by_t = (y_t^{(1)}, y_t^{(2)} ,\ldots,
y_t^{(c)})$ where 
$y_t^{(j)} = x_t^{(j)} + g_t^{(j)}$ and $g_t^{(j)}$ is an AWGN  with zero mean and variance $\sigma^{2}=N_{0}/2$.
%
%
%

Using the same notation as in Sect.~\ref{sec:ldpc_dec}, the pipelined BP decoding algorithm 
applying to LDPCCC is illustrated in Algorithm~\ref{alg:ldpccc_dec}.
Same as the LDPC decoding algorithm, the LDPCCC decoding algorithm consists of 4 main procedures: initialization, horizontal step,
vertical step and making hard decisions. Moreover, for each of these procedures, multiple threads can be used in executing the computations in parallel and all the threads will follow the same instructions with no divergence occurring, except when making hard decisions.
}

\begin{algorithm}[t]
\caption{\color{black} BP Decoding Algorithm for LDPCCC}
\label{alg:ldpccc_dec}
\begin{algorithmic}[1]                    
{\color{black} 
\State Set time $t=1$
\Statex [Initialization]
\State Shift $c$ new variable nodes (denoted by $\bv_t$) together with 
their channel messages $\mu_n$ into the first processor.
\For{$i = 1, 2, ... , I - 1$}
	\If{$t \geq i(m_s + 1)$}
			\State Shift the variables $\bv_{t-i(m_s+1)}$ along with their associated
			variable-to-check messages $\beta$'s from the $i$-th processor to the $(i + 1)$-th processor.
	\EndIf
\EndFor
}{\color{black} 
\For{Processor $i$, $i = 1, 2, ... , I$}
	\Statex [Horizontal step]
	\State Update the $(c - b)$ check nodes corresponding to the $t-(i-1)(m_s+1)$-th 
	block row of $\bH_{[0,\infty]}$ (as in
\eqref{eq:ldpcccH}) using
		\begin{equation}
		\alpha_{mn}=2\tanh^{-1}\Bigg(\prod_{n\prime\in\mathcal{N}(m)\backslash n}\tanh\left(\frac{\beta_{mn\prime}}{2}\right)\Bigg)
		\label{eq:ldpccc_bph}
		\end{equation}
	\Statex [Vertical step]
	\State Update the variable nodes $\bv_{t-(i-1)(m_s+1)-m_s}$ using
		\begin{equation}
		\beta_{mn}=\mu_{n}+\sum_{m\prime\in\mathcal{M}(n)\backslash m}\alpha_{m\prime n}
		\label{eq:ldpccc_bpvs}
		\end{equation}
\EndFor
}
{\color{black} 
	\Statex [Making hard decision for the variable nodes leaving the last processor]}
	\State {\color{black} Evaluate the \emph{a posteriori} LLRs of the frame 
	$\bv_{t-(I-1)(m_s+1)-m_s}$ using
		\begin{equation}
		\beta_n = \mu_n + \sum_{m'\in \mathcal{M}(n)}\alpha_{m'n}
		\end{equation}
}
	\State {\color{black} Make hard decisions based on the LLRs 
		\begin{eqnarray}
		\hat{v}_{n} & = & \begin{cases}
		0 & {\rm if}\ \beta_{n} \ge0 \\
		1 & {\rm if}\ \beta_{n} <0
		\end{cases}
		\label{eq:ldpccc_hrddcs}
		\end{eqnarray}
}		%
\State {\color{black} Set time $t=t+1$ and Go to [Initialization] }
%
%
\end{algorithmic}
\end{algorithm}

\section{Graphics Processing Unit and CUDA Programming}
\label{sec:cuda}

\begin{figure}[t]
\centering
\includegraphics[width=0.9\columnwidth]{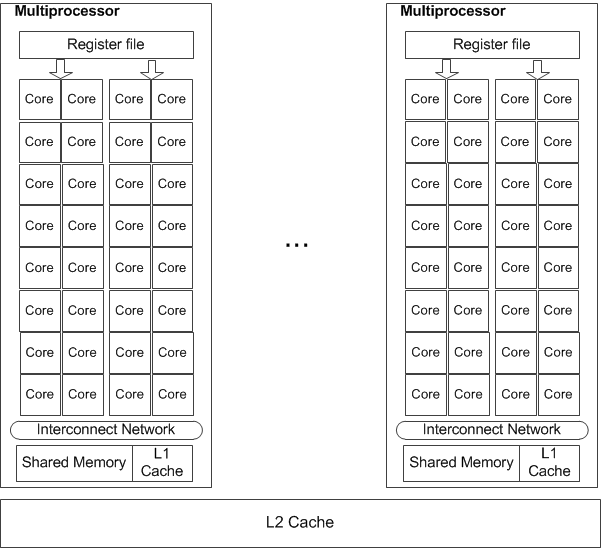} 
\caption{Hardware architecture of a typical GPU. Each large rectangle 
	denotes a streaming multiprocessor (SM) and each small square
in a SM represents a stream processor or core.}
\label{Flo:arcgpu} 
\end{figure}

%
%
A graphics processing unit (GPU) consists of multi-threaded, multi-core
processors. GPUs follow the single-instruction multiple-data (SIMD) paradigm.
That is to say, given a set of data (regarded as a stream), the same operation or
function is applied to each element in the stream by different processing units
in the GPUs simultaneously. Figure~\ref{Flo:arcgpu} shows a simplified
architecture of the latest GPU device. It contains a number of multiprocessors
called streaming multiprocessors (or SMs). Each SM contains a group of stream
processors or cores and several types of memory including registers, on-chip
memory,
 L2 cache and the most
plentiful dynamic random-access memory (DRAM). The L1 cache is dedicated to each
multiprocessor and the L2 cache is shared by all multiprocessors. Both caches are
used to cache accesses to local or global memory. The on-chip memory has a small
capacity (tens of KB) but it has a low latency~\cite{Nvidia2011}.
 
In our work, 
{\color{black} the GPU used is a GTX460, }
which has 7 SMs and 768 MB global memory.
Each SM contains 48 cores \cite{Nvidia2009}. Moreover, the 64 KB on-chip memory
is configured as 48 KB shared memory and 16 KB L1 cache for each SM because the
more shared memory is utilized, the better.

%
%
CUDA (Compute Unified Device Architecture) is a parallel computing architecture
developed by Nvidia. 
%
%
%
In a CUDA program, computations are performed as a sequence of functions called
parallel kernels. Each kernel
is typically invoked on a massive number of threads. 
Threads are first grouped into thread blocks and blocks are further grouped into 
a grid.
A thread block contains a set of concurrently executing threads, and the size of
all blocks are the same with an upper limit $\Omega_{{\rm max}}$. 
{\color{black} In current GPUs with compute capability 2.x, }
$\Omega_{\max}=1024$. 

%
%
In an abstract level, the CUDA devices use different memory spaces, which have
different characteristics.
These memory spaces includes global memory, local memory, shared memory, constant
memory, texture memory, and registers.
The global memory and texture memory are the most plentiful 
but have the largest access latency followed by constant memory, registers, and
shared memory. 

%
%
CUDA's hierarchy of threads map to a hierarchy of processors on the GPU. A GPU
executes one or more kernel grids and a SM executes one or more thread blocks.
{\color{black} In current GPUs with compute capability 2.x, }
the SM creates, manages, schedules, and executes
threads in groups of 32 parallel threads called \emph{warps}. A warp is the
execution unit and executes one common instruction at a time. So full efficiency
is realized when threads of a warp take the same execution path.

%
%
{\color{black} In CUDA programming, 
the first important consideration is the coalescing global memory accesses. }
Global memory resides in the device memory and is accessed via 32-, 64-, or
128-byte memory transactions. These memory transactions must be naturally
aligned (i.e. the first address is a multiple of their size). 

When a warp executes an instruction that accesses the global memory, it coalesces
the memory accesses of the threads within the warp into one or more of these
memory transactions depending on the size of the word accessed by each thread and
the distribution of the memory addresses across the threads. 
%
%

\section{Implementation of Decoders for LDPC Codes and LDPCCCs}
\label{sec:dec}

\subsection{GPU-based LDPC Decoder}


We implement our decoders using the standard BP decoding algorithm. 
According to the CUDA programming model, the granularity of a thread execution 
and a coalesced memory access is a warp. Full efficiency is realized when all 
threads in a warp take the same execution path
and the coalesced memory access requirement is satisfied.
Thus, we propose to decode $\Gamma$ codewords simultaneously, where $\Gamma$ is
an integer multiple of a warp (i.e., multiple of 32). For each decoding cycle,
$\Gamma$ codewords will be input, decoded, and ouput together and in parallel. 




Recall that an LDPC code can be represented by its parity-check matrix or a
Tanner graph. 
A non-zero element in the parity-check
matrix corresponds to an edge in the Tanner graph. 

In the LDPC decoder, messages are bound to the edges in the Tanner graph (or the
$1$'s in the parity-check matrix $\bH$). So we store the messages according to
the positions of $1$'s. Besides, the channel messages corresponding to the
variable nodes are required. To reuse the notation, we denote the data structure
storing the messages between the variable nodes and the check nodes as $\bH$
while the the data structure storing the channel messages as $\bV$. The
difficulty of the CUDA memory arrangement lies on the fact that for practical
LDPC codes with good performance, the positions of the $1$'s are scattered in the
parity-check matrix. 

\begin{figure}[t]
\centering
\includegraphics[width=0.9\columnwidth]{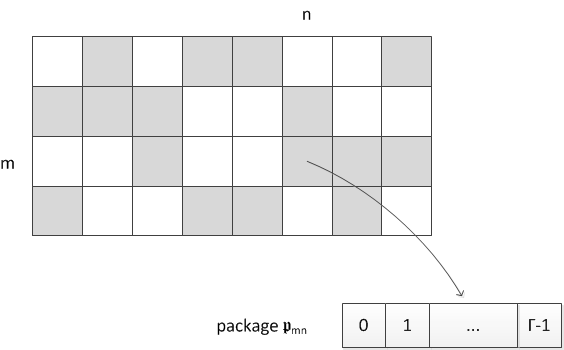}
\caption{Data Structure for an LDPC code. Each gray square denotes a non-zero entry in the 
parity-check matrix. $\Gamma$ is a multiple of a warp (i.e., multiple of 32).}
\label{fig:datastruct_pack}
\end{figure}

%
%
First, in the BP decoding procedure, although there are two kinds of messages,
namely, the variable-to-check messages and the check-to-variable messages, at
every step of the iteration, only one kind of message is needed to be stored,
i.e., after the check-node updating step, only the check-to-variable messages
$\alpha$'s are stored in the $\bH$ and after the variable-node updating step,
only the variable-to-check messages $\beta$'s are stored in the $\bH$. Second, in
our new decoder architecture, $\Gamma$ ($\Gamma$ is a multiple of a warp)
codewords are decoded together and hence the messages of $\Gamma$ codewords are
also processed together. We number the distinct codewords as $0, 1, ...,
\Gamma-1$ and we use the same notations for the messages as before, i.e.,
$\beta_{mn}(\gamma)$ is the message from variable node $n$ to check node $m$
corresponding to the $\gamma$-th codeword and $\alpha_{mn}(\gamma)$ is the
message from check node $m$ to variable node $n$ corresponding to the $\gamma$-th
codeword. Since all the $\Gamma$ codewords messages share the same Tanner graph,
messages of the $\Gamma$ distinct codewords corresponding to the same edge can be
grouped into one package and stored linearly. Let $\spp_{mn}$ denote the package
corresponding to the edge connecting variable node $n$ and check node $m$. Then
in package $\spp_{mn}$, $\beta_{mn}(0), \beta_{mn}(1), ..., \beta_{mn}(\Gamma-1)$
or $\alpha_{mn}(0), \alpha_{mn}(1), ..., \alpha_{mn}(\Gamma-1)$ are stored
contiguously. This is shown in Figure~\ref{fig:datastruct_pack}. Different
packages $\spp_{mn}$'s are aligned linearly according to their corresponding
positions in the parity-check matrix --- row-by-row, and left to right for each
row. That implies the messages associated to one check node are stored
contiguously. 

\begin{figure*}[t]
\begin{align}
& \qquad\qquad\text{variable node } n \nonumber \\
&\text{\qquad 0\qquad 1\qquad 2\qquad 3\qquad\ 4\qquad\ 5\qquad\ 6\qquad\ 7} \nonumber \\
\mathbf{H} = &
 \begin{bmatrix}
  0 & 1_{(0)} & 0 & 1_{(1)} & 1_{(2)} & 0 & 0 & 1_{(3)} \\
  1_{(4)} & 1_{(5)} & 1_{(6)} & 0 & 0 & 1_{(7)} & 0 & 0 \\
  0 & 0 & 1_{(8)} & 0 & 0 & 1_{(9)} & 1_{(10)} & 1_{(11)} \\
  1_{(12)} & 0 & 0 & 1_{(13)} & 1_{(14)} & 0 & 1_{(15)} & 0
 \end{bmatrix}
\begin{array}{c}
0\\1\\2\\3
\end{array}
\text{check node } m
\label{eq:simpleH}
\end{align}
\end{figure*}

\begin{rem}
To be consistent with the use of memory locations in computer programming, all the indices
 of the data structures 
in this paper starts from $0$. 
\end{rem}


The advantage of this arrangement is obvious. Since $\Gamma$ is a multiple of 32,
the memory segment for every package is naturally aligned when the data type
belongs to one of the required data types (i.e., with word size of 1-, 2-, 4-, or
8-byte). In addition, the structure of the parity-check matrix $\bH$ is shared by
the $\Gamma$ codewords. As these $\Gamma$ data elements are processed together,
they can be accessed by $\Gamma$ contiguous threads and hence the global memory
is always accessed in a coalesced way. 
{\color{black} We also ensure that the threads within a warp always follow the same execution path with no
divergence occurring (except when making hard decisions on the received bits). 
Then both the memory access and the thread execution are optimal
and efficient. }

%
%
We also need to store the details of the parity-check matrix. Two lookup tables
denoted by $LUT_c$ and $LUT_v$ will be kept. $LUT_c$ is used in the check-node
updating process and $LUT_v$ is used in the variable-node updating process. The
two tables store the indices of the data accessed in the two updating processes
and both are two-dimensional. The first dimension is to distinguish different
check nodes, i.e., $LUT_c[m]$ is associated with the $m$-th check node or the
$m$-th row. Each $LUT_c[m]$ records the indices
of the messages related to the $m$-th check node. The two lookup tables are
shared by all $\Gamma$ codewords.

\begin{align}
LUT_c = 
\begin{bmatrix}
0 & 1 & 2 & 3 \\
4 & 5 & 6 & 7 \\
8 & 9 & 10& 11 \\
12& 13& 14& 15 \\
\end{bmatrix}
\begin{array}{c}
0\\1\\2\\3
\end{array}
\text{check node } m
\label{eq:LUT_c}
\end{align}

\begin{align}
LUT_v = 
\begin{bmatrix}
4 & 12 \\
0 & 5 \\
6 & 8 \\
1 & 13 \\
2 & 14 \\
7 & 9 \\
10 & 15 \\
3 & 11
\end{bmatrix}
\begin{array}{c}
0\\1\\2\\3\\4\\5\\6\\7
\end{array}
\text{variable node } n
\label{eq:LUT_v}
\end{align}

\begin{exm}
Consider the parity-check matrix in \eqref{eq:simpleH}. The corresponding
data structure begins with the package $\spp_{01}$, which is followed by
$\spp_{03}$, $\spp_{04}$, $\spp_{07}$, $\spp_{10}$, $\spp_{11}$, ...,
$\spp_{30}$, ..., $\spp_{36}$. The subscripts of the nonzero
entries indicate the sequences (or positions) of the associated
data in the entire data structure, starting from $0$. The $LUT_c$ and $LUT_v$ are
shown in \eqref{eq:LUT_c} and \eqref{eq:LUT_v}.

It is seen that the
size of $LUT_c$ can be reduced by only storing the first address of the data in
each row, namely, $LUT_c[0]$ only store $0$, $LUT_c[1]$ only store $4$ and so on
for all $m = 0,1, ..., M-1$. Particularly, for regular LDPC codes with a unique
row weight $d_c$, the indices in $LUT_c[m]$ for the $m$-th check node are from
$m\cdot d_c$ to $m\cdot d_c+d_c-1$. As for the $LUT_v$, the indices are normally
irregular and random. Hence a full-indexed lookup table is required for $LUT_v$.

The $LUT_c$ and $LUT_v$ lookup tables are stored in the constant or
texture memory in the CUDA device so as to be cached to reduce the access time.
\end{exm}

\begin{figure}[t]
\centering
\includegraphics[width=1\columnwidth]{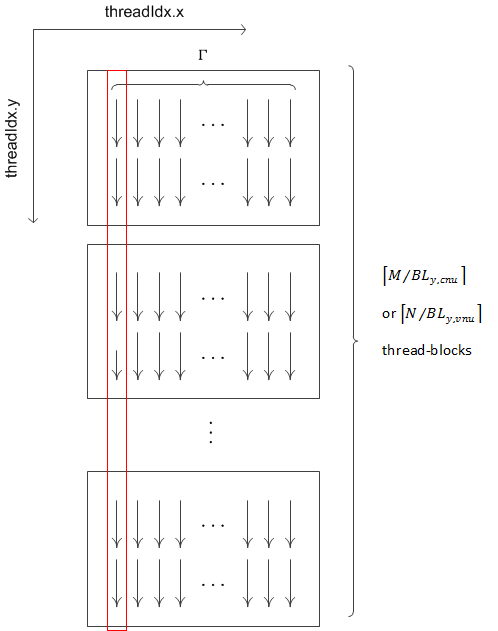}
\caption{Two dimensional thread layout of the check-node/variable-node updating kernel.}
\label{fig:thread_layout_bp_pack}
\end{figure}

%
%
A separate thread is assigned to process each check node or
each variable node in the updating kernel. 
Hence, $\Gamma$ threads can be assigned
to process the data of $\Gamma$ codewords simultaneously. So, 
a two dimensional thread hierarchy is launched. The first dimension
is for identifying the different codewords while the second dimension is for
processing different check nodes or variable nodes. The thread layout is
illustrated in Fig.~\ref{fig:thread_layout_bp_pack}. For each thread block, we
allocate $\Gamma$ threads in the threadIdx.x dimension\footnote{In CUDA, threads
are linear in the threadIdx.x dimension.}, and $BL_y$ threads in the threadIdx.y
dimension. Each thread-block contains $BL_y\times\Gamma$ threads, which
should be within the thread-block size limit (1024 for the current device). 
The total number of thread-blocks is
determined by the number of check nodes $M$ or the number of variable nodes $N$.
We denote $BL_y$ in the check-node updating kernel as $BL_{y,cnu}$ and the one in
the variable-node updating kernel as $BL_{y,vnu}$. Then the numbers of thread
blocks are given by $\lceil M/BL_{y,cnu}\rceil$ and $\lceil N/BL_{y,vnu}\rceil$,
respectively. In Fig.~\ref{fig:thread_layout_bp_pack}, the threads marked by the vertical
rectangular are processing the same codeword.

In the check-node updating kernel and the variable-node updating kernel, the
forward-and-backward calculation is adopted as in \cite{kschischang_factor_2001}. 
The shared memory is used to cache the involved data so as to avoid
re-accessing the global memory. Due to the limited size of the shared memory, the
size of the thread-block should not be too large. Consider a $(J,L)$ QC-LDPC
code. For each check node, there are $2L$ data elements to be stored. Denote the
shared memory size by $Z_{{\rm shared}}$ and the size of each data by $Z_{{\rm
data}}$. Consequently in the check-node updating kernel, the thread-block size,
denoted by $\Omega_{{\rm cnu}}$, is limited by
\begin{equation}
\Omega_{{\rm cnu}}\leq\frac{Z_{{\rm shared}}}{2LZ_{{\rm data}}}.
\label{eq:bp_th_limit}
\end{equation}
In addition, $\Omega_{{\rm cnu}} = BL_{y,cnu}\times\Gamma$ and 
$\Omega_{\rm vnu}=BL_{y,vnu}\times\Gamma$.

With such a thread layout, the threads access the memory in a straightforward
pattern. For example, for the check-node updating kernel, a two-dimensional
thread hierarchy with a total size of $\lceil \frac{M}{BL_{y,cnu}}\rceil\times
BL_{y,cnu}\times\Gamma$ is launched. During the memory access, every $\Gamma$ threads 
are one-to-one mapped to $\Gamma$ data in a message package. 
Hence, coalesced memory access is guaranteed. 

\subsection{GPU-based LDPCCC Decoder}

The decoding algorithm and the pipelined LDPCCC decoder architecture have been
introduced in Section \ref{sec:ldpccc_dec}. The LDPCCCs studied in our work
are derived from QC-LDPC codes as described in Section \ref{sec:ldpccc_derive}.
So our LDPCCC decoder is confined to the LDPCCCs with the parity-check matrix
$\bH_{[0,\infty]}$ of this kind of structure.

\subsubsection{Data Structure}\label{sect:data}

\begin{figure}[t]
\centering
\includegraphics[width=\columnwidth]{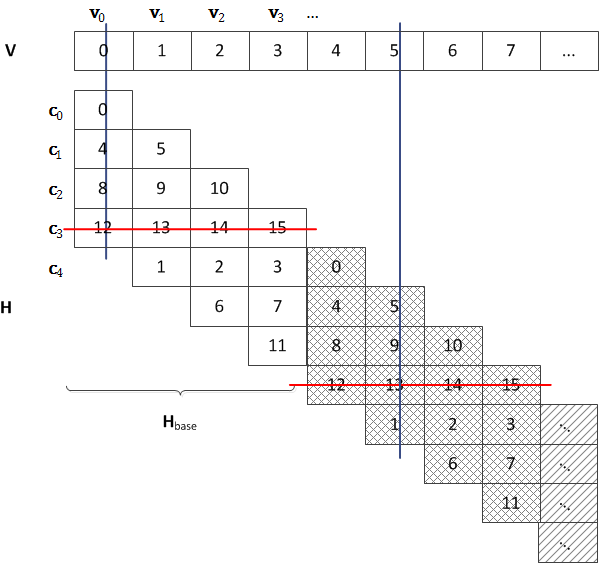}
\caption{The periodic structure of the parity-check matrix of the LDPCCCs.}
\label{fig:ldpccc_str_re}
\end{figure}

The LDPC convolutional codes are decoded continuously. We will thus refer to an
LDPCC code sequence $\bv_{[0,\infty]}$ = $[\bv_0, \bv_1,\ldots,\bv_{\infty}]$ as a
\emph{code stream} and $\bv_i$, $i=0,1,\ldots,\infty$ as a \emph{code frame} or
\emph{variable frame}. A code stream is constrained with the parity-check
matrix $\bH_{[0,\infty]}$ by
\[
\bv_{[0,\infty]}\bH_{[0,\infty]}^T=\b0.
\]
The parity-check matrix of the LDPCCC is shown in Figure~\ref{fig:ldpccc_str_re}. 
It is seen that the check nodes are grouped into layers. Each variable-node frame 
is connected to $m_s+1$ ($4$ here) check layers in the parity-check matrix. 
Let $c$ denote the size of $\bv_i$, $i=0,1,...,\infty$ and $c-b$ denote the size 
of each check layer. Thus the code rate is $b/c$.
 
We will use the same notations as in Section \ref{sec:ldpccc_derive}. The LDPCCC
is derived from a $(J, L)$ QC-LDPC base code $\bH_{QC}$ which has $J\times L$
sub-matrices and the size of each sub-matrix is $p\times p$. $\bH_{QC}$ is first
divided into $\Lambda\times\Lambda$ sub-blocks\footnote{Note that a ``sub-block''
is different from a ``sub-matrix''.} ($\Lambda=4$ in
Figure~\ref{fig:ldpccc_str_re}) and each sub-block contains several sub-matrices.
We have $c=L/\Lambda\times p$ and $c-b=J/\Lambda\times p$. Referring to Section
\ref{sec:ldpccc_derive}, we denote the unwrapped parity-check matrix of the
QC-LDPC code as
\[
\centering
\bH_{base} = 
\begin{bmatrix}
\bH_{QC}^{L} \\
\bH_{QC}^{U} 
\end{bmatrix}.
\]
The $\bH_{[0,\infty]}$ of the derived LDPCCC is a repetition of  $\bH_{base}$.
Denotingthe number of edges in $\bH_{base}$ by  $E$, we have $E=J\times L\times p$.
%

%
%
In designing the LDPCCC decoder, the first thing to consider is the amount of memory required 
to store the messages. Like the LDPC decoder, we store the messages according to the edges 
in the parity-check matrix. Let $I$ denote the number of iterations in the LDPCCC decoding. Then
$I$ processors are required in the pipelined decoder.  Although the parity-check matrix of the 
LDPCCC is semi-infinite, the decoder only needs to allocate memory for $I$ processors. 
Hence the total size of the memory required for storing the messages passing between the variable 
nodes and check nodes is $I\times E$ units. And the total size of the memory required for storing 
these channel messages is $I\times c$.

Next, we will describe the hierarchical data structure for the LDPCCC decoder memory space.
To reuse the notation, we use $\bH$ to denote the memory space for the messages on the 
edges and $\bV$ to denote the memory space for the channel messages.
The $\bH$ is a multi-dimensional array with two hierarchies. 
First, we divide the entire memory space into $I$ groups corresponding to the $I$ processors
and we use the first hierarchy of $\bH$ as the data structure for each group. That is
$\bH[i]$, $i=0,1,...,I-1$ denote the data structure for the $I$ processors, respectively.
Second, recall that the parity-check matrix in 
Figure~\ref{fig:ldpccc_str_re} is derived from $\bH_{base}$ which is divided into $16$ 
non-zero sub-blocks and each sub-block has a size of 
$(pJ/\Lambda)\times(pL/\Lambda)$. Thus in each group, $\bH[i]$ is 
also divided into $16$ sub-blocks, denoted by the second hierarchy of $\bH$, namely,
 $\bH[i][j]$, where $j=0, 1,...,15$. Every $\bH[i][j]$ stores the messages associated 
with one sub-block. On the other hand, the memory for the channel messages 
is simpler: $\bV[i]$, $i=0,1,...,I\cdot(m_s+1)-1$ will be allocated.
Finally, to optimize the thread execution and memory access,
$\Gamma$ LDPC convolutional code streams are decoded simultaneously, 
where $\Gamma$ is a multiple of a warp. 
Thus every $\Gamma$ data are combined into one package and take up one memory unit. 

%
%
An LDPCCC decoder uses the BP algorithm to update the check nodes and variable
nodes. The BP decoding procedures are based on the parity-check matrix
$\bH_{[0,\infty]}$. With the data structure to store the messages, the decoder
also needs the structure information of $\bH_{[0,\infty]}$ for understanding the
connections between the check nodes and the variable nodes. This information can
be used to calculate the index of the data being accessed during the updating.
Due to the periodic property of the constructed LDPCCC, the
structure of $\bH_{base}$ is shared by all the processors. 
We label the $16$ sub-blocks in $\bH_{base}$ with the numbers $0,1,\ldots,15$.

In addition, in the decoder, the $I$ check-node layers or $I$ variable-node frames being
updated simultaneously in the $I$ processors are separated by an interval of
$m_s+1$. Since $\bH_{[0,\infty]}$ also has a period of $T = m_s+1$, at any time
slot, the $I$ processors require the same structure information in updating the
check nodes or the variable nodes, as seen in Figure~\ref{fig:ldpccc_str_re}. The
lookup tables used in check-node updating and variable-node updating are denoted
as $LUT_c$ and $LUT_v$, respectively. The two lookup tables will then store the
labels of the sub-blocks in $\bH_{base}$ that are involved in the updating process. Besides,
another lookup table $LUT_{sub}$ will be used to store the ``shift
numbers\footnote{For a QC-LDPC base matrix, the information is the ``shift
number'' of each $p\times p$ sub-matrix ($-1$ represents the all-zero matrix, $0$
represents the identity matrix, $l$ represents cyclically right-shifting the
identity matrix $l$ times).}'' of the sub-matrices in each sub-block. 

\begin{exm}
The $LUT_c$ and $LUT_v$ for the LDPCCC in Figure~\ref{fig:ldpccc_str_re} are
\begin{equation} 
LUT_c = 
\begin{bmatrix}
1 & 2 & 3 & 0 \\
6 & 7 & 4 & 5 \\
11& 8 & 9 & 10 \\
12& 13& 14& 15
\end{bmatrix}
\label{eq:lut_c_ldpccc}
\end{equation}
and
\begin{equation} 
LUT_v = 
\begin{bmatrix}
0 & 4 & 8 & 12 \\
5 & 9 & 13& 1 \\
10& 14& 2 & 6 \\
15&  3& 7 & 11
\end{bmatrix}.
\label{eq:lut_v_ldpccc}
\end{equation}
\end{exm}

\subsubsection{Decoding Procedures}
\label{sub:decproc_ldpccc}

Based on the discussion in Section \ref{sec:ldpccc_dec}, the detailed decoding procedures are
 as follows.

\begin{enumerate}
\item At time slot $0$, the first code frame $\bv_0$ enters Processor $1$. This
means the corresponding memory space $\bV[0]$ will be filled with the channel
messages of $\bv_0$. Then the channel messages will be propagated to the
corresponding check nodes. Hence, referring to Fig.~\ref{fig:ldpccc_str_re} and
\eqref{eq:lut_v_ldpccc}, $\bH[0][0]$, $\bH[0][4]$, $\bH[0][8]$ and $\bH[0][12]$
will be filled with the same channel messages $\bV[0]$.

Next, the first check layer of $\bv_0$, i.e., $\bc_0$, will be updated based on
the messages from $\bv_0$, namely, the messages stored in $\bH[0][0]$ (they are
the only messages available to $\bc_0$). 

\item At time slot $1$, the second code frame $\bv_1$ enters Processor $1$. Hence
the memory space $\bV[1]$, $\bH[0][5]$, $\bH[0][9]$, $\bH[0][13]$ and $\bH[0][1]$
will be filled with the messages of $\bv_1$. Then, the check layer $\bc_1$ are
updated in a similar way as the check layer $\bc_0$. However, both the messages
from $\bv_0$ and $\bv_1$, i.e., messages stored in $\bH[0][4]$ and $\bH[0][5]$,
are used in the updating of $\bc_1$ based on the index information in $LUT_c[1]$.
The procedure at time slot $1$ is shown in Figure~\ref{fig:slot2}.

The procedure goes on. When $\bv_3$ has been input and check layer $\bc_3$ has
been updated, all the check-to-variable messages needed to update the variable
layer $\bv_0$ are available. So $\bv_0$ will be updated with the channel messages
in $\bV[0]$ and the check-to-variable messages in $\bH[0][0]$, $\bH[0][4]$,
$\bH[0][8]$, and $\bH[0][12]$. 
Now, $\bv_0$ is at the end of Processor $1$ and is about to be shifted
to Processor $2$. Instead of copying the memory from one location to another, all
we need to do is to specify that the memory $\bv_0$ ``belongs'' to Processor $2$.

\item At the next time slot, i.e., time slot $4$ (time slot $m_s+1$), the new
code frame $\bv_4$ comes. The messages will be stored in $\bv[4]$ and
$\bH[1][0]$, $\bH[1][4]$, $\bH[1][8]$ and $\bH[1][12]$. Now there are two check
layers to update, $\bc_0$ and $\bc_4$. It is noted that $\bc_4$ are updated based
on all the available messages in $\bH[0][1]$, $\bH[0][2]$, $\bH[0][3]$, and
$\bH[1][0]$ while $\bc_0$ are updated based on the updated messages only in
$\bH[0][0]$. This insufficient updating of check nodes only occurs to the first
$m_s$ code frames. After the updating of the check nodes, the code frame $\bv_1$
is at the end of Processor $1$ and will be updated. There is no code frame
arriving at the end of Processor $2$ yet.

\item At time slot $I\cdot(m_s+1)-1$, the entire memory space of $\bV$ and $\bH$
are filled with messages. 
$\bv_0$ and its
associated messages are at the end of Processor $I$ (as being labeled) while
$\bv_{i\cdot(m_s+1)-1}$ is the latest code frame input into Processor $1$. Next,
the check nodes in the $I$ check layers of the $I$ processors will be updated in
parallel. After the updating of the check nodes, all the variable nodes which are
leaving Processor $i$ ($i=1,2,...,I$) are updated. Specifically, the variable
nodes $\bv_{i\cdot(m_s+1)-1}$, $i=1,2,\ldots,I$ are to be updated. Furthermore,
$\bv_0$ is about to leave the decoder. Hard decision will be made based on the
\emph{a\ posteriori} LLR of $\bv_0$. Then the memory space of $\bv_0$, $\bV[0]$,
$\bH[0][0]$, $\bH[0][4]$, $\bH[0][8]$ and $\bH[0][12]$ are cleared for reuse. At
the next time slot $I\cdot(m_s+1)$, the new code frame $\bv_{I\cdot(m_s+1)}$
comes in the decoder and these memory space will be filled with the messages of
$\bv_{I\cdot(m_s+1)}$.
\begin{rem}
In our GPU-based decoder, all the check nodes (variable nodes) needed to be updated in the $I$ processors are
processed in parallel by multiple threads.
\end{rem}
\item Note that the LDPCCC matrix has a period of $T=m_s+1$ ($4$ here). Hence, at
time slot $t\ge I\cdot(m_s+1)$, $\bv_t$ enters the decoder and reuses the memory
space of $\bv_{\tau}$ where $\tau=t \mod (I\cdot(m_s+1))$. Furthermore, we let
$\kappa=t \mod (m_s+1)$. Then the check layer $\bc_{\kappa+(i-1)\cdot(m_s+1)}$ in
Processor $i$ ($i=1,2,\ldots,I$) will be updated followed by the updating of the
code frame $\bv_{\kappa+i\cdot(m_s+1)-m_s}$. Moreover, the ``oldest'' code frame
residing in the decoder --- $\bv_{t-I\cdot(m_s+1)}$ --- is about to leave the
decoder and hard decisions will be made on it.

So the entire LDPCCC decoder possesses a circulant structure, as shown in
Figure~\ref{fig:ldpccc_cir_dec}. The memory is not shifted except for the one
associated with the code frame which is leaving the decoder. Instead, the
processor are ``moving'' by changing the processor label of each code frame.
Correspondingly, the ``entrance'' and ``exit'' are moving along the data
structure. This circulant structure reduces the time for memory manipulation and
simplifies the decoder.

\subsubsection{Parallel Thread Hierarchy}

As described in Sect.~\ref{sect:data}, the memory associated with each entry in the
$\bH$ matrix is a message package containing $\Gamma$ messages from $\Gamma$ code
streams. So there is a straightforward mapping between the thread hierarchy and
the data structure. In the
check-node-updating kernel (or variable-updating-kernel), a two dimensional
thread hierarchy of size $I\cdot(c-b)\times \Gamma$ (or $I\cdot c \times \Gamma$)
is launched, where $(c-b)$ (or $c$) is mapped to the total number of check nodes
(or variable nodes) being updated in $I$ processors. The size of one of the
dimensions (i.e., $\Gamma$) is mapped to the number of code streams. Like in
LDPC decoder, $\Gamma$ will be configured as the
\emph{threadIdx.x} dimension and $(c-b)$ (or $c$) will be the \emph{threadIdx.y}
dimension in the CUDA thread hierarchy. The $\Gamma$ threads in the
\emph{threadIdx.x} dimension is contiguous and will access the $\Gamma$ data in
each message package for coalesced access.

\begin{figure}[t]
\centering
\subfloat[Updating at time slot 1.]{
\includegraphics[width=0.35\columnwidth]{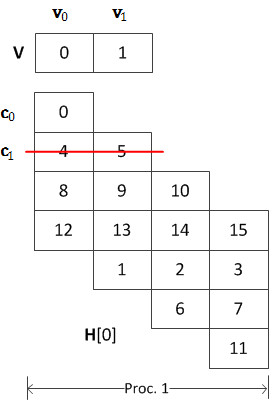}
\label{fig:slot2}
}
\hfill
\subfloat[Updating at time slot 4.]{
\includegraphics[width=0.5\columnwidth]{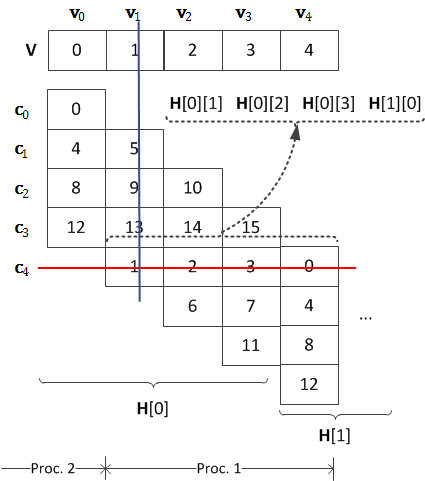}
\label{fig:slot5}
}
\caption{Illustration of the procedures of a LPDCCC decoder.
The horizontal line denotes the updating of the row.
The vertical line denotes the updating of a column.}
\label{fig:ldpccc_dec_proc}
\end{figure}
\begin{figure}[t]
\centering
\includegraphics[width=0.9\columnwidth]{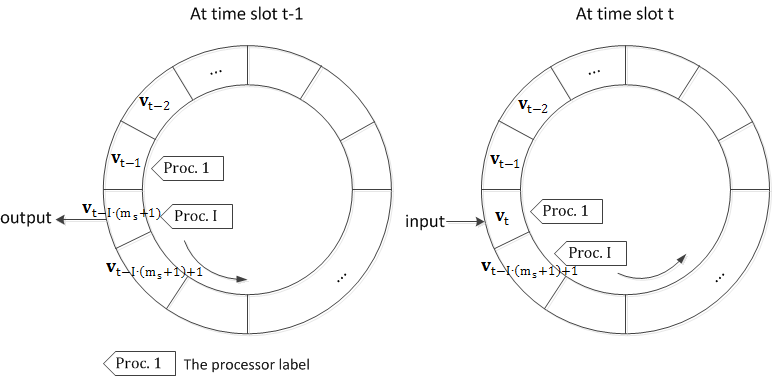}
\caption{Illustration of the circulant structure of LPDCCC decoder.}
\label{fig:ldpccc_cir_dec}
\end{figure}

\end{enumerate}

{\color{black}
\subsection{CPU-based LDPC and LDPCCC Decoders}
We implement both the serial CPU-based LDPC decoder and LDPCCC decoder
using the C language. 
As CPUs with multiple cores are very common nowadays, we further 
 implement a multi-thread CPU-based LDPCCC decoder using OpenMP. 
OpenMP \cite{chandra2001parallel} is a portable, 
scalable programming interface for shared-memory parallel computers. 
It can be used to explicitly direct multi-threaded, shared memory 
parallelism. A straightforward application of the OpenMP is to parallize the intensive 
loop-based code with the \emph{\#pragma omp parallel for} 
directive. Then the executing threads will be automatically allocated 
to different cores on a multi-core CPU.

The horizontal step and the vertical step in 
 Algorithm~\ref{alg:ldpccc_dec} involve intensive computing. 
On a single-core CPU, the updating of the different nodes are processed with 
a serial \emph{for} loop. Since the updating of different nodes can be performed independent
of one another, it is ideal to 
parallelize the \emph{for} loop with the \emph{\#pragma omp parallel for} 
directive in the OpenMP execution on a multicore CPU. Hence, in our implementation, 
we issue multiple threads to both the updating of the check nodes 
\eqref{eq:ldpccc_bph} and the updating of the variable nodes 
\eqref{eq:ldpccc_bpvs} in the multi-thread CPU-based LDPCCC decoder.
}

\section{Results and Discussion}
\label{sec:results}

\begin{table}[t]
\centering
\begin{tabular}{
>{\centering}p{0.2\columnwidth}>{\centering}p{0.27\columnwidth}>{\centering}p{0.33\columnwidth}}
\toprule 
 & \textbf{CPU}  & \textbf{GPU}\tabularnewline
\midrule
\midrule 
Platform & Intel Xeon  & Nvidia GTX460\tabularnewline
\midrule 
Number of cores & 4   & 7 $\times$ 48 = 336\tabularnewline
\midrule 
Clock rate & 2.26 GHz  & 0.81 GHz\tabularnewline
\midrule 
Memory & 8 GB DDR3 RAM  & 768 MB global memory and 48 KB shared memory \tabularnewline
\midrule 
{\color{black} Maximum number of threads} & {\color{black} $8$}  & {\color{black}---} \tabularnewline
\midrule 
Maximum thread-block size & ---  & $1024$ threads \tabularnewline\midrule 
{\color{black} Programming language } & {\color{black}C/OpenMP}  & {\color{black}  CUDA C} \tabularnewline
\bottomrule
\end{tabular}
\caption{Simulation environments.}
\label{Flo:expenviro}
\end{table}

\begin{table}[t]
\centering
\begin{tabular}{
>{\centering}p{0.1\columnwidth}>{\centering}p{0.1\columnwidth}
>{\centering}p{0.1\columnwidth}>{\centering}p{0.2\columnwidth}
>{\centering}p{0.2\columnwidth}}
\toprule 
\textbf{Code} & \textbf{$J\times L$} & $p$ & $c\times(c-b)$  & \textbf{Number of Edges} \tabularnewline
\midrule
\midrule 
A & $4\times24$ & $422$ & $2532\times422$ & $40512$\tabularnewline
\midrule 
B & $4\times24$ & $632$ & $3792\times632$ & $60672$\tabularnewline
\midrule 
C & $4\times24$ & $768$ & $4608\times768$ & $73728$\tabularnewline
\midrule 
D & $4\times24$ & $1024$ & $6144\times1024$ & $98304$\tabularnewline
\bottomrule
\end{tabular}
\caption{Parity-check matrices of the QC-LDPC codes used in the LDPC decoder. 
They are also used to derive the LDPCCCs A' to D'.}
\label{Flo:codechar_ldpccc}
\end{table}

\subsection{The Experimental Environment}
{\color{black}  The CPU being used is an Intel Xeon containing $4$ cores. Moreover, it can handle up to $8$ threads at a time. 
The serial CPU-based decoders are developed using C and 
the multi-threaded CPU-based LDPCCC decoder is developed using OpenMP.
Note that for the serial CPU-based decoders, only one of 
the $4$ cores in the CPU will be utilized. 
The GPU used in this paper is a GTX460 containing $336$ cores and the GPU-based
decoders are developed using CUDA C. 
Furthermore, in our simulations, $32$ codewords are decoded simultaneously  in the GPU decoders, i.e., $\Gamma=32$.
 }
Details of the CPU and GPU used in our simulations are
presented in Table~\ref{Flo:expenviro}. 

Table~\ref{Flo:codechar_ldpccc} shows the
characteristics of the QC-LDPC codes under test. For Code A to code D, $J=4$ and
$L=24$ thus giving the same code rate of $(24-4)/24=5/6$. 
These codes are further used to derived regular LDPCCCs.
In order to avoid confusion, we denote the derived LDPCCCs as Code A' to Code D'.
It can be readily shown that the $(3, 4, 24)$-LDPCCCs A' to D' 
have the same code rate of 
$5/6$.

\begin{rem}
Note that although QC-LDPC codes are adopted in the simulation, the new GPU-based
LDPC decoder is able to decode other LDPC codes like randomly-constructed regular
or irregular codes.
\end{rem}

\subsection{The Decoding Time Comparison}

{\color{black} 
In order to optimize the speed and to minimize the data transfer
between the CPU (host) and the GPU (device), we generate and process the data, 
including the codeword and the AWGN noise, directly 
on the GPU. 
After hard decisions have been made on the received bits, 
the decoded bits are transferred to the CPU which counts the  
number of error bits. Since the data transfer occurs only at the end of the iterative 
decoding process, the transfer time (overhead) is almost negligible compared with
time spent in the whole decoding process. }

In the following, we fix the number of decoding iterations and the simulation
terminates after $100$ block/frame errors are received. By recording the total 
number of blocks/frames decoded and the total time taken\footnote{\color{black} In the case of 
the GPU-based decoders, the total
time taken includes the GPU computation time, the time spent in transferring data between the CPU and GPU, etc.
However, as explained above, the GPU computation time dominates the total time
while the overhead is very small.}, we can compute the
average time taken to decode one block/frame.

\begin{table}[t]
\centering
\begin{tabular}{
c|
>{\centering}p{0.07\columnwidth}>{\centering}p{0.07\columnwidth}
>{\centering}p{0.07\columnwidth}|
>{\centering}p{0.07\columnwidth}>{\centering}p{0.07\columnwidth}
>{\centering}p{0.07\columnwidth}|
>{\centering}p{0.10\columnwidth}
}
\toprule
\textbf{Code} 
& $C_{\rm GPU}$ & $T_{\rm GPU}$ (s) & $t_{\rm GPU}$ (ms) 
& $C_{\rm CPU}$ & $T_{\rm CPU}$ (s) & $t_{\rm CPU}$ (ms) & Speedup ($\frac{t_{\rm CPU}}{t_{\rm GPU}}$) \tabularnewline
\midrule
\midrule
A & 2832  & 6  & 2.12 & 4058  & 1270 & 313  & 148 \tabularnewline
\midrule                      
B & 12768 & 37 & 2.9  & 11664 & 5350 & 458  & 158 \tabularnewline
\midrule                      
C & 21664 & 74 & 3.4  & 20046 & 10950 & 546 & 161 \tabularnewline
\midrule                    
D & 82624 & 371 & 4.5 & 70843 & 51580 & 728 & 162 \tabularnewline
\bottomrule
\end{tabular}
\caption{Decoding time for the GPU-based LDPC decoder and the serial CPU-based decoder at
$E_b/N_0$=3.2 dB. 30 iterations are used. $C$ represents the total number of decoded
codewords; $T$ denotes the total simulation time and $t$ is the average simulation
time per codeword.}
\label{tbl:newdectime}
\end{table}

\subsubsection{LDPC decoders}
The GPU-based
decoder  and the serial CPU-based decoder are tested with 30 iterations at a $E_b/N_0$ of 3.2 dB. 
Table~\ref{tbl:newdectime} shows the number of transmitted codewords and the
simulation times for different codes. 

We consider the average time for decoding one codeword for the serial CPU-based decoder,
i.e., $t_{{\rm CPU}}$. We observe that $t_{{\rm CPU}}$ increases from Code A to
Code D due to an increasing number of edges in the codeword.
Further, we consider the average time for decoding one codeword for the GPU-based
decoder, i.e., $t_{{\rm GPU}}$. Similar to the serial CPU-based decoder, $t_{{\rm GPU}}$
increases from Code A to Code D. 

Finally, we compare the simulation times of the serial CPU-based decoder and the
GPU-based decoders by taking the ratio $t_{{\rm CPU}}/t_{{\rm GPU}}$. The results
in Table~\ref{tbl:newdectime} indicate that the GPU-based decoder accomplishes 
speedup improvements from $148$ times to $162$ times 
compared with the serial CPU-based decoder.

\subsubsection{LDPCCC decoders}
{\color{black}
We decode the LDPC convolutional codes A' to D' at a 
$E_b/N_0$ of $3.1$ dB with $I = 20$. 
First, we show the average decoding times for Code A' and Code C' 
when different numbers of threads are used in the CPU-based decoders. 
The results are shown in Table~\ref{tbl:dectime_openmp}. The serial CPU-based
decoder corresponds to the case with a single thread. 
We observe that the decoding time is approximately inversely proportional to the number of threads used --- up to $4$ threads. However, the time does not improve much when the number of threads increases to $6$ or $8$. The reason is as follows. The CPU being used has $4$ cores, which can execute
up to $4$ tasks in fully parallel. Hence, compared with using a single thread, there is an almost $4$ times improvement when $4$ threads are used. As the number of threads increases beyond $4$, however, all the threads cannot 
really be executed at the same time by the $4$ cores. Consequently, further time improvement is small when more than $4$ threads are used. 
}

\begin{table}[t]
\centering
\begin{tabular}{cccccc}
\toprule
\multirow{2}{*}{\textbf{Code}} & \multicolumn{5}{c}{\textbf{ Number of threads used}}
\tabularnewline
\cmidrule{2-6}
& 1 & 2 & 4 & 6 & 8 \tabularnewline
\midrule
\midrule
A' & 39 & 20 & 11 & 10 & 9   \tabularnewline
\midrule
C' & 73 & 38 & 21 & 19 & 17 \tabularnewline
\bottomrule
\end{tabular}
\caption{\color{black} Average decoding time (ms) per code frame for the quad-core CPU-based decoder when different numbers of threads are used.}
\label{tbl:dectime_openmp}
\end{table}

{\color{black}
Next, we compare the decoding times of the LDPCCC decoders when 
GPU-based and CPU-based decoders are used to decode Code A' to Code D'.
For the CPU-based decoders, we consider the cases where 
a single thread and $8$ threads are used, respectively.
Table~\ref{tbl:dectime_ldpccc} shows the results. 
As explained above, limited by the number of cores ($4$ only)
in the CPU, the CPU-based decoder can only improve the speed 
by about $4$ times even when the number of threads increases from $1$ to $8$. 
We also observe that compared with the serial CPU-based decoder,
 the GPU-based LDPCCC 
decoder can achieve $170$ to $200$ times speedup improvement.
Compared with the $8$-thread CPU-based decoder, the GPU-based LDPCCC 
decoder can also accomplish $39$ to $46$ times speedup improvement.
}
%

\begin{table*}[t]
\centering
\begin{tabular}{
>{\centering}p{0.07\columnwidth}|
>{\centering}p{0.07\columnwidth}>{\centering}p{0.07\columnwidth}>{\centering}p{0.07\columnwidth}|
>{\centering}p{0.1\columnwidth}>{\centering}p{0.1\columnwidth}>{\centering}p{0.1\columnwidth}
>{\centering}p{0.1\columnwidth}>{\centering}p{0.1\columnwidth}|
>{\centering}p{0.1\columnwidth}>{\centering}p{0.1\columnwidth}>{\centering}p{0.1\columnwidth}
}
\toprule
\textbf{Code} 
& $C_{\rm GPU}$ & $T_{\rm GPU}$ (s) & $t_{\rm GPU}$ (ms)
& $C_{\rm CPU}$ & $T_{\rm CPU-1}$ (s) & $t_{\rm CPU-1}$ (ms) 
& $T_{\rm CPU-8}$ (s) & $t_{\rm CPU-8}$ (ms) 
& $\frac{t_{\rm CPU-1}}{t_{\rm CPU-8}}$ 
& $\frac{t_{\rm CPU-1}}{t_{\rm GPU}}$ 
& $\frac{t_{\rm CPU-8}}{t_{\rm GPU}}$ 
 \tabularnewline
\midrule
\midrule
A & 3136 & 0.73 & 0.23 & 2846 & 112 & 39 & 28 & 9 & 4.3 & 170 & 39 \tabularnewline
\midrule
B & 6272 & 1.95 & 0.31 & 5716 & 345 & 60 & 79 & 14 & 4.3 & 194 & 45  \tabularnewline
\midrule
C & 14400 & 5.4 & 0.38 & 13303 & 976 & 73 & 230 & 17 & 4.3 & 192 & 45  \tabularnewline
\midrule
D & 43680 & 21.0 & 0.48 & 37451 & 3590 & 96 & 834 & 22 & 4.4 & 200 & 46  \tabularnewline
\bottomrule
\end{tabular}
\caption{\color{black} Decoding time for the GPU-based LDPCCC decoder and the CPU-based decoders at
$E_b/N_0$=3.1 dB. $I=20$ processors are used. $C$ represents the total number of decoded
frames; $T$ denotes the total simulation time and $t$ is the average simulation
time per frame. CPU$-1$ and CPU$-8$ denote the use of $1$ thread and $8$ threads, respectively, in the  CPU-based decoder.}
\label{tbl:dectime_ldpccc}
\end{table*}

\section{Conclusion}
\label{sec:conclude}

In this paper, efficient decoders for LDPC codes
and LDPC convolutional codes based on the GPU parallel architecture are implemented. 
%
%
%
By using efficient data structure and thread layout, the thread divergence is 
minimized and the memory can be accessed in a coalesced way.
All decoders are flexible and scalable. First, they can
decode different codes by changing the parameters. Hence, the programs need very
little modification. Second, they should be to run on the latest or even future
generations of GPUs which possess more hardware resources.
{\color{black} For example, if there are more cores/memory in the GPU, we can readily decode more codes, say  $\Gamma=64$ codes as compared with $\Gamma=32$ codes used in this paper,  at the same time. }
 These are actually
advantages of GPU parallel architecture compared to other parallel solutions
including FPGA or VLSI.
{\color{black}  We will report our results in the future when we have the opportunity to run our proposed mechanism in other GPU families. }

Compared with the traditional serial CPU-based decoders, results show that the proposed
GPU-based decoders can achieve $100\times$ to $200\times$ speedup. The actual
time depends on the particular codes being simulated. 
{\color{black}  When compared with the $8$-thread CPU-based decoder, the GPU-based  
decoder can also accomplish $39$ to $46$ times speedup improvement.
Thus the simulation time can be reduced from months to weeks or days when a GPU-based decoder is used.
In summary, our results show that the proposed
GPU-based LDPC/LDPCCC decoder has obvious advantages in the decoding time compared with CPU-based decoders. }




\bibliographystyle{IEEEtran}
\bibliography{parallelbib-new}





\end{document}